\newif\ifpdf
\ifx\pdfoutput\undefined
   \pdffalse
\else
   \pdfoutput=1
   \pdftrue
\fi

\ifpdf
\documentclass[runningheads, pdftex]{cl2emult}

\else
\documentclass[runningheads, dvips]{cl2emult}
\fi

\usepackage{amsmath}
\usepackage{amssymb}
\usepackage{graphicx} 

\ifpdf
    \usepackage{color}
    \definecolor{myred}{rgb}{0.5,0,0}
    \definecolor{myblue}{rgb}{0,0,0.75}
    \definecolor{mygreen}{rgb}{0,0.5,0}
   \usepackage[pdftex,
               pdftitle={Calculating Concentration-Sensitive Capital Charges
with Conditional Value-at-Risk},
               pdfsubject={Credit portfolio risk},
               pdfkeywords={One-factor model; capital charge;
quantile derivative; Expected Shortfall},
               pdfauthor={Dirk Tasche, Ursula Theiler},
               pdfstartview=FitH,
               breaklinks=true,
               bookmarks=false,
               colorlinks=true,
               citecolor=myred,
               linkcolor=myblue,
               urlcolor=mygreen]{hyperref}
\else
   \usepackage{hyperref}
\fi

\begin{document}
\renewcommand{\thefootnote}{\fnsymbol{footnote}}

\title*{Calculating Concentration-Sensitive Capital Charges
with Conditional Value-at-Risk}
\toctitle{Calculating Concentration-Sensitive Capital Charges
with Conditional Value-at-Risk}
%
%
\titlerunning{Calculating Concentration-Sensitive Capital Charges}
%
\author{Dirk Tasche\inst{1} \and Ursula Theiler\inst{2}}
\authorrunning{Dirk Tasche and Ursula Theiler}
%
%
\institute{Deutsche Bundesbank, Postfach 10 06 02, 60006 Frankfurt am Main,
Germany\newline E-mail: tasche@ma.tum.de\newline The opinions expressed in
this note are those of the authors and do not necessarily reflect views shared
by the Deutsche Bundesbank or its staff. \and Risk Training,
Carl-Zeiss-Str.~11,
     83052 Bruckm\"uhl,
     Germany\newline E-mail: theiler@risk-training.org}

\maketitle              

\renewcommand{\thefootnote}{\arabic{footnote}}
\setcounter{footnote}{0}

\begin{abstract}
By mid 2004, the Basel Committee on Banking Supervision (BCBS) is expected to
launch its final recommendations on minimum capital requirements in the
banking industry. Although there is the intention to arrive at capital charges
which concur with economic intuition, the risk weight formulas proposed by the
committee will lack an adequate treatment of concentration risks in credit
portfolios. The question arises whether this problem can be solved without
recourse to fully-fledged portfolio models. Since recent practical experience
shows that the risk measure Conditional Value-at-Risk (CVaR) is particularly
well suited for detecting concentrations, we develop the semi-asymptotic
approach by Emmer and Tasche in the CVaR context and compare it with the
capital charges recently suggested by the Basel Committee. Both approaches are
based on the same Vasicek one-factor model.
\end{abstract}

\section{Introduction}

From an economic point of view, the risks that arise in a portfolio need to be
covered by a corresponding amount of capital that is applied as a cushion to
absorb potential losses. The question of calculating the risk contributions of
single assets in a portfolio corresponds to the problem of calculating capital
charges to cover occurring loss risks. This allocation issue can be considered
as well from a regulatory as well as from an internal perspective, leading to
capital charges per asset by the regulatory and the economic capital,
respectively.

Considering the different regulations for credit risk, we observe that in the current
regulatory regime (Basel I Accord) almost no risk adjustment can be identified. By
mid 2004, the BCBS is expected to launch
its final recommendations on new minimum capital requirements in the banking
industry (Basel II Accord). Although there is the intention to arrive at
improved more risk sensitive capital charges of the credit risk bearing
assets, the risk weight formulas proposed by the committee will lack an
adequate treatment of concentration risks in credit portfolios as we will show
below. The main problem is that the Basle II model assumes that all credits
are of equal, infinitesimal small exposure, i.e. that the credit portfolio is
infinitely fine grained. However, in real world portfolios, this basic
assumption does not hold.

From an internal perspective, banks are putting high efforts into the
development of internal credit risk models that allow the risk measurement of
the portfolio credit risk. Comprehensive research work has been done to
develop methods of how to calculate risk contributions in an appropriate way
(see e.g.\ \cite{Denault}, \cite{kalk}, \cite{PBR}, \cite{Tasche99},
\cite{Theiler03}). However, from a practical point of view, the allocation
problem, i.e. the question of how the single assets contribute to the overall
portfolio risk, cannot yet be considered as solved. Risks are broken down in
different ways, taking into account correlation effects and concentration risk
to different extents (see e.g.\ \cite{BC03b}). Banks that are using
internal credit risk models in most cases need enormous calculation efforts to
estimate the overall portfolio risk and the risk contributions of single
assets and sub-portfolios.

In this paper, we give a survey on an approach to
calculate risk contributions of single assets in an analytical way that avoids
calculation intensive simulation efforts. This \emph{semi-asymptotic} approach
slightly extends the Basel II approach and takes into account concentration
effects. Thus, it can be viewed as a bridging to relate regulatory and
internal risk measurement. Additionally, it is based on the new risk measure
of Conditional Value at Risk that has been proven to be appropriate for bank
wide loss risk measurement (see for instance
\cite{AT02}, \cite{RU02}, \cite{Theiler03}).

The paper is organized as follows. In Section~\ref{Basel} we briefly
introduce the capital charges as they are suggested by the Basel Committee.
Section~\ref{semi} presents the semi-asymptotic approach to capital charges in
case of CVaR as risk measure. In Section~\ref{num} we illustrate the both
approaches with a numerical example.

\section{The Basel II Model}
\label{Basel}

We give a short presentation of the reasoning that lead to the current
suggestions by the Basel Committee. In particular, we introduce the so-called
Vasicek one-factor model that was originally proposed in \cite{Gordy01} for
use in the forthcoming rules on capital charges.

We consider  a portfolio loss variable $L_n$ that is defined by
\begin{equation}
  \label{eq:8}
  L_n \, =\, L_n(u_1, \ldots, u_n)\,=\, \sum_{i=1}^n u_i\,
\mathbf{1}_{\{\sqrt{\rho_i}\,X+\sqrt{1-\rho_i}\,\xi_i\le c_i\}},
\end{equation}
where $u_i\ge 0$, $i = 1, \ldots, n$,
denotes the weight or the exposure of asset $i$ in the portfolio, $0 < \rho_i
< 1$ and $c_i \ge 0$, $i = 1, \ldots, n$, are constants, and $X, \xi_1,
\ldots, \xi_n$ are independent random variables with continuous distributions.
The constants $c_i$ are called \emph{default thresholds}. They have to be
calibrated in order to fix the probabilities of default of the assets. The
random variable $X$ is interpreted as the change of an economic factor that
influences all the assets in the portfolio but to different extents. The
so-called \emph{asset correlation} $\rho_i$ measures the degree of the $i$-th
asset's exposure to the systematic risk expressed by $X$. The random variables
$\xi_i$ are assumed to model the idiosyncratic (or specific) risk of the
assets.

Equation (\ref{eq:8}) implies the following representation for the conditional
variance of the loss $L_n$ given $X$
\begin{equation}
  \label{eq:9a}
  \mathrm{var}[L_n\,|\,X=x]\,=\,\sum_{i=1}^n u_i^2\,\mathrm{P}
\bigl[\xi_i\le
\textstyle{\frac{c_i-\sqrt{\rho_i}\,x}{\sqrt{1-\rho_i}}}\bigr]\,
\bigl(1-\mathrm{P}
\bigl[\xi_i\le
\textstyle{\frac{c_i-\sqrt{\rho_i}\,x}{\sqrt{1-\rho_i}}}\bigr]\bigr).
\end{equation}
We assume that the probabilities of default are not too small and that the
correlations with the economic factor are not too high, i.e.\ in precise terms
that
$\inf_i c_i > -\infty$ and $\sup_i \rho_i < 1$.
Then from (\ref{eq:9a}) it follows that in case of independent, identically
distributed $\xi_1, \xi_2, \ldots$  we have \addtocounter{footnote}{1}
\begin{equation}
\lim_{n\to\infty} \mathrm{E}\bigl[\mathrm{var}[L_n\,|\,X]\bigr] = 0\quad
\text{if and only
if\footnotemark[\thefootnote]}\quad
\lim_{n\to\infty} \sum_{i=1}^n u_i^2  =
0.\label{eq:cond}
\end{equation}
\footnotetext[\thefootnote]{%
Of course, here we admit an additional dependence of $u_i$ on $n$, i.e.\ $u_i%
= u_{i,n}$.}%
Since
\begin{equation}
  \label{eq:var}
    \mathrm{var}[L_n]\,=\, \mathrm{E}\bigl[\mathrm{var}[L_n
\,|\,X]\bigr] + \mathrm{var}\bigl[\mathrm{E}[L_n\,|\,X]\bigr],
\end{equation}
the conditional expectation $\mathrm{E}[L_n\,|\,X]$ appears to be a natural
approximation of $L_n$ as soon as (\ref{eq:cond}) is fulfilled, i.e.\ as soon
as the concentrations in the portfolio are not too big. Indeed, the
approximation
\begin{equation}
  \label{eq:basel}
  L_n \,\approx \, \mathrm{E}[L_n\,|\,X]
\end{equation}
is fundamental for the Basel~II approach to credit risk capital charges.
For $\alpha\in (0,1)$ and any random variable $Y$, define the
$\alpha$-quantile (or the Value-at-Risk (VaR)) of $Y$ by
\begin{equation}
  \label{eq:4}
  q_\alpha(Y)\,=\, \text{VaR}_\alpha(Y) \,=\,
\inf\bigl\{ y\in\mathbb{R}: \mathrm{P}[Y\le y]\ge\alpha\bigr\}.
\end{equation}
Note that
\begin{equation}
  \label{eq:quant}
\mathrm{E}[L_n\,|\,X=x]\,=\,\sum_{i=1}^n u_i\,\mathrm{P}\bigl[\xi_i\le
\textstyle{\frac{c_i-\sqrt{\rho_i}\,x}{\sqrt{1-\rho_i}}}].
\end{equation}
Since the right-hand side of (\ref{eq:quant}) is a decreasing function in $x$,
one then deduces from (\ref{eq:basel}) that
\begin{subequations}
\begin{equation}
  \label{eq:approx}
  q_\alpha(L_n) \,\approx\, \sum_{i=1}^n u_i\,\mathrm{P}\bigl[\xi_i\le
\textstyle{\frac{c_i-\sqrt{\rho_i}\,q_{1-\alpha}(X)}{\sqrt{1-\rho_i}}}\bigr].
\end{equation}
Assuming that the $\xi_i$ are all standard normally distributed then yields
\begin{equation}
  \label{eq:normal}
 q_\alpha(L_n) \,\approx\, \sum_{i=1}^n u_i\,\Phi\bigl(
\textstyle{\frac{c_i-\sqrt{\rho_i}\,q_{1-\alpha}(X)}{\sqrt{1-\rho_i}}}\bigr),
\end{equation}
\end{subequations}
where $\Phi$ denotes the standard normal distribution function. The linearity
of the right-hand side of (\ref{eq:normal})
in the vector $(u_1, \ldots, u_n)$ suggests the choice of
\begin{equation}
  \label{eq:charge}
\text{Basel II charge}(i) \,=\, u_i\,\Phi\bigl(
\textstyle{\frac{c_i-\sqrt{\rho_i}\,q_{1-\alpha}(X)}{\sqrt{1-\rho_i}}}\bigr)
\end{equation}
as the capital requirement of asset $i$ in the portfolio with the loss
variable $L_n$. Up to an adjustment for the maturity of the loan which can be
neglected in the context of this paper, (\ref{eq:charge}) is just the form
of the risk weight functions that was provided by the BCBS in \cite{BC03}.

\section{Calculating Risk Contributions with the Semi-Asymptotic Approach}
\label{semi}
\subsection{Definition of semi-asymptotic capital charges}
In the following we review the approach by \cite{EmmerTasche} for the
definition of the risk contributions of single credit assets if risk is
measured with Value-at-Risk (or just a quantile at fixed level). However, we
consider here the risk measure Conditional Value-at-Risk which turned out to
be more attractive from a conceptual point of view. We call our approach
semi-asymptotic because, in contrast to Basel~II where all exposures are
assumed to be infinitely small, we keep one exposure fixed and let the others
tend to infinitely small size.

We consider here a special case of (\ref{eq:8}) where $\rho_1 = \tau$, $c_1 =
a$ but $\rho_i = \rho$ and $c_i = c$ for $i > 1$, and $\sum_{i=1}^n u_i = 1$.
Additionally, we assume that $u_1 = u$ is a constant for all $n$ but that
$u_2, u_3, \ldots$ fulfills (\ref{eq:cond}).

In this case, the portfolio loss can be represented by
\begin{multline}
  \label{eq:3.0}
  L_n(u, u_2, \ldots, u_n) \,=\, u\,\mathbf{1}_{\{\sqrt{\tau}\,X +
  \sqrt{1-\tau}\,\xi \le a\}} + \\
(1-u) \sum_{i=2}^n u_i\,
  \mathrm{1}_{\{\sqrt{\rho}\,X + \sqrt{1-\rho}\,\xi_i\le c\}},
\end{multline}
with $\sum_{i=2}^n u_i = 1$. Transition to the limit for $n\to \infty$ in
(\ref{eq:3.0}) leads to the \emph{semi-asymptotic} percentage loss function
\begin{equation}
  \label{eq:3.1}
  L(u)\,=\,u\,\mathbf{1}_D + (1-u)\,Y
\end{equation}
with $D = \{\sqrt{\tau}\,X +
  \sqrt{1-\tau}\,\xi \le a\}$ and $Y = \mathrm{P}\bigl[\xi \le \frac{c-
  \sqrt{\rho}\,x}{\sqrt{1-\rho}}\bigr]\Big|_{x=X}$. Of course, a natural
  choice for $\tau$ might be $\tau = \rho$, the mean portfolio asset
  correlation.
For $\alpha\in (0,1)$ and any random variable $Z$, define the
the Conditional Value-at-Risk (CVaR) (or Expected Shortfall, see \cite{AT02})
at level $\alpha$ of $Z$ by
\begin{subequations}
\begin{equation}
  \label{eq:CVaR}
\mathrm{CVaR}_\alpha(Z) \,=\,\mathrm{E}\bigl[Z\,|\,Z \ge q_\alpha(Z)\bigr].
\end{equation}
As by (\ref{eq:3.1}) we have
\begin{multline}
  \label{eq:decomposition}
\mathrm{CVaR}_\alpha(L(u)) \,=\, u\,\mathrm{P}[D\,|\,L(u) \ge q_\alpha(L(u))]\\
+ (1-u)\, \mathrm{E}\bigl[Y\,|\,L(u) \ge q_\alpha(L(u))\bigr],
\end{multline}
\end{subequations}
the following definition is rather near at hand.\\[1ex]
\emph{
The quantity
\begin{equation}\label{de:1}
u\,\mathrm{P}[D\,|\,L(u) \ge q_\alpha(L(u))]
\end{equation}
is called\/ \emph{semi-asymptotic CVaR capital charge} (at level $\alpha$) of the
loan with exposure $u$ (as percentage of total portfolio exposure) and default
event $D$ as in (\ref{eq:3.1}).
}\\[1ex]
The capital charges we suggest in  Definition (\ref{de:1}) have to be
calculated separately, i.e.\ for each asset an own model of type
(\ref{eq:3.1}) has to be regarded. This corresponds to a bottom-up approach
since the total capital requirement for the portfolio is determined by adding
up all the capital charges of the assets. Note that the capital charges of
Definition (\ref{de:1}) are not portfolio invariant in the sense of
\cite{Gordy01}. However, in contrast to the portfolio invariant charges, the
semi-asymptotic charges take into account not only correlation but also
concentration effects. In particular, their dependence on the exposure $u$ is
not merely linear since also the factor $\mathrm{P}[D\,|\,L(u) \ge
q_\alpha(L(u))]$ depends upon~$u$. Definition (\ref{de:1}) is in line with the
general definition of risk contributions (cf.\ \cite{L96}, \cite{Tasche99})
since (\ref{eq:3.1}) can be considered a two-assets portfolio model.

\subsection{Calculation of semi-asymptotic capital charges}
If $F_0$ and $F_1$ denote the conditional distribution functions of $Y$ given
$\mathbf{1}_D = 0$ and $\mathbf{1}_D = 1$ respectively, the distribution
function of $L(u)$ is given by
\begin{equation}\label{eq:3.2a}
  \mathrm{P}[L(u) \le z] \, =\, p\,F_1\bigl(\frac{z-u}{1-u}\bigl) +
  (1-p)\,F_0\bigl(\frac{z}{1-u}\bigl),
\end{equation}
where $p = \mathrm{P}[D]$ is the \emph{default probability} of the loan under
consideration. By means of (\ref{eq:3.2a}), the quantile
  $q_\alpha(L(u))$ can be numerically computed. For the conditional
  probability which is part of Definition (\ref{de:1}), we obtain
  \begin{equation}
    \label{eq:3.3}
 \mathrm{P}[D\,|\,L(u) \ge z]\,=\,
\frac{p\,\bigl(1 - F_1(\frac{z-u}{1-u}\bigr)\bigr)}{\mathrm{P}[L(u) \ge z]}.
  \end{equation}
Denote by
$\Phi_2(\cdot, \cdot;\theta)$ the distribution function of the bivariate standard
normal distribution with correlation $\theta$. If we assume that $X$ and $\xi$
are independent and both standard normally distributed, we obtain $p =
\Phi(a)$, and can derive for the
conditional distribution functions from (\ref{eq:3.2a}) that
\begin{subequations}
\begin{align}\label{eq:F1}
  F_1(z) &=
  \begin{cases}
     1 - p^{-1}\, \Phi_2\bigl(a, \frac{c -
    \sqrt{1-\rho}\,\Phi^{-1}(z)}{\sqrt{\rho}}; \sqrt{\tau}\bigr), & z
\in (0,1) \\
0, & \mbox{otherwise,}
  \end{cases}\\
\intertext{and}
  F_0(z) &=
  \begin{cases}
    (1- p)^{-1}\, \Phi_2\bigl(- a,- \frac{c -
    \sqrt{1-\rho}\,\Phi^{-1}(z)}{\sqrt{\rho}}; \sqrt{\tau}\bigr), & z
\in (0,1) \\
0, & \mbox{otherwise.}
  \end{cases}\label{eq:F0}
\end{align}
\end{subequations}
Let, similarly to the case of (\ref{eq:F1}) and (\ref{eq:F0}),
$\Phi_3(\cdot, \cdot, \cdot;\Sigma)$ denote the distribution function of the
tri-variate standard normal distribution with correlation matrix $\Sigma$.
Define the function $g$ by
\begin{subequations}
  \begin{align}
    g(x, \beta, z) & = \frac{x-\sqrt{1-\beta}\; \Phi^{-1}(z)}{\sqrt{\beta}}
\intertext{and the correlation matrix $\Sigma_{\rho,\tau}$ by}
\Sigma_{\rho,\tau} &= \textstyle\begin{pmatrix} 1 & \sqrt{\rho\,\tau} & \sqrt{\rho}\\
 \sqrt{\rho\,\tau} & 1 & \sqrt{\tau}\\
\sqrt{\rho} &  \sqrt{\tau} & 1\end{pmatrix}.
  \end{align}
Then, in order to arrive at $\mathrm{CVaR}_\alpha(L(u))$,
the conditional expectation of $Y$ given $\{L(u) \ge z\}$ from
(\ref{eq:decomposition}) can
be calculated according to
\begin{multline}
  \label{eq:ey}
\mathrm{E}\bigl[Y\,|\,L(u) \ge z\bigr] \,=\,
\mathrm{P}[L(u)\ge z]^{-1} \Big( \Phi_3\bigl(c, a, g(c,\rho,\textstyle{\frac{z-u}{1-u})};
\Sigma_{\rho, \tau}\bigr) + \\
\Phi_2\bigl(c, g(c,\rho,\textstyle{\frac{z}{1-u})};
\sqrt{\rho}\bigr) - \Phi_3\bigl(c, a, g(c,\rho,\textstyle{\frac{z}{1-u})};
\Sigma_{\rho, \tau}\bigr)\Big).
\end{multline}
\end{subequations}

\section{Numerical Example}
\label{num}
\setcounter{figure}{0}

We illustrate the previous results by a numerical
example. In our focus is a portfolio that is driven by systematic
risk only (the variable $Y$ in (\ref{eq:3.1})) and enlarge this portfolio with
an additional loan (the indicator $\mathbf{1}_D$ in (\ref{eq:3.1})).

In our example, the portfolio modeled by $Y$ has a quite moderate credit
standing which is expressed by its expected loss $\mathrm{E}[Y] = 0.025 =
\Phi(c)$. By choosing $\rho = 0.1$ as asset correlation we arrive at a
portfolio with a rather strong exposure to systematic risk. For the sake of
simplicity we choose $\tau = \rho$, i.e.\ the exposure to systematic risk of
the additional loan is identical to the exposure of the existing portfolio.
However, we assume that the additional loan enjoys a quite high
credit-worthiness as we set $p = \mathrm{P}[D] = 0.002 = \Phi(a)$.
\refstepcounter{figure}
\ifpdf
\begin{figure}[ht]
  \centering
  \resizebox{\height}{10.0cm}{\includegraphics[width=10.0cm]{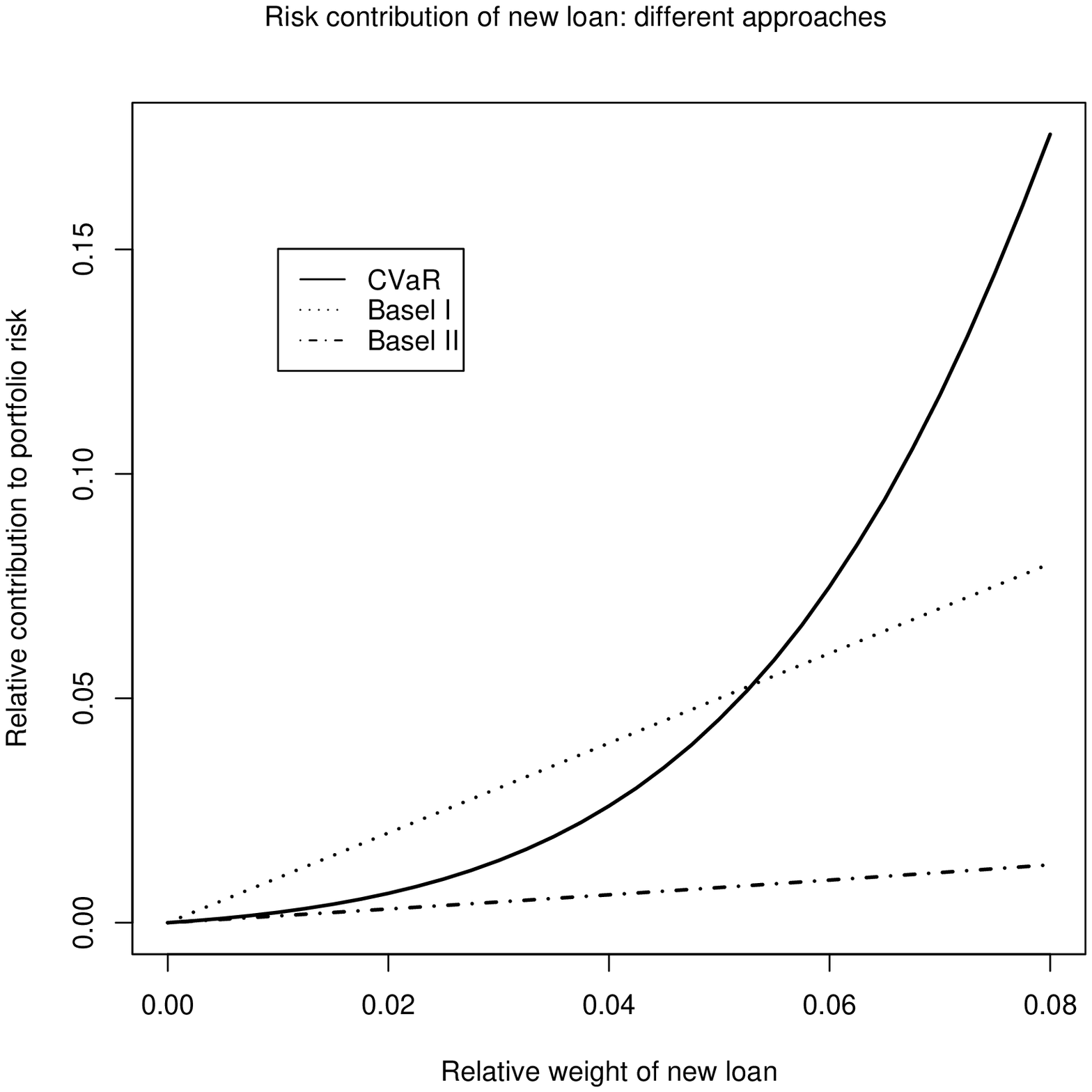}}
  \parbox{10.0cm}{\footnotesize{}Figure \thefigure:
Relative risk contribution of new loan as function of the relative weight of
  the new loan. Comparison of contribution to true Conditional Value-at-Risk (CVaR),
and the contributions according to
  the Basel~II and Basel~I Accords.}\label{fig:2}
\end{figure}
\else
\begin{figure}[ht]
  \centering
\resizebox{\height}{10.0cm}{
\includegraphics[width=10.0cm]{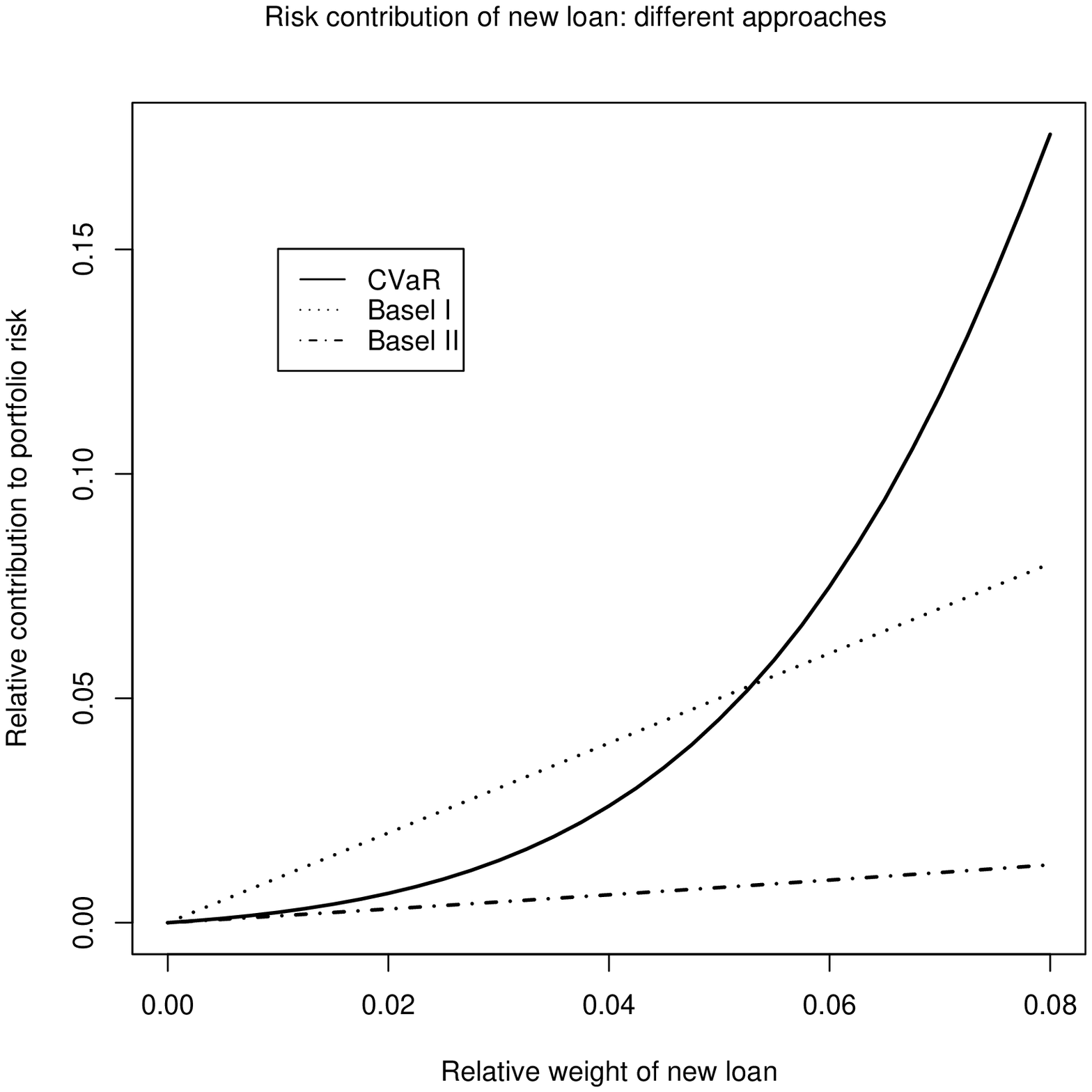}}
  \parbox{10.0cm}{\footnotesize{}Figure \thefigure:
Relative risk contribution of new loan as function of the relative weight of
  the new loan.
Comparison of contribution to true Conditional Value-at-Risk (CVaR),
and the contributions according to
  the Basel~II and Basel~I Accords.}\label{fig:2}
\end{figure}
\fi
Figure~\ref{fig:2} illustrates the relative contribution of the new loan to
the risk of the portfolio loss variable $L(u)$.  The contribution is expressed
as a function of the relative weight $u$ of the new loan in the portfolio and
calculated according to three different methods. The first of the depicted
methods relates to the relative contribution to true portfolio CVaR at level
$\alpha = 99.9\%$, defined as the ratio of the contribution to CVaR according
to Definition~(\ref{de:1}) and portfolio CVaR, i.e.\ the function
\begin{equation}\label{eq:me1}
  u\,\mapsto\,\frac{u\,\mathrm{P}[D\,|\,L(u) \ge
  q_{\alpha}(L(u))]}{\mathrm{CVaR}_{\alpha}(L(u))},
\end{equation}
where the conditional probability has to be evaluated by means of
(\ref{eq:3.3}) and (\ref{eq:F1}). In the denominator, $\mathrm{CVaR}$ is
calculated according to (\ref{eq:decomposition}) and (\ref{eq:ey}).
Moreover, curves are drawn for the
Basel~II approach, i.e.\ the function
\begin{equation}\label{eq:contribs}
  u \, \mapsto\,
  \frac{u\,\Phi\bigl(\frac{a-\sqrt{\tau}\,q_{1-\alpha}(X)}{\sqrt{1-\tau}}\bigr)}
  {u\,\Phi\bigl(\frac{a-\sqrt{\tau}\,q_{1-\alpha}(X)}{\sqrt{1-\tau}}\bigr) +
  (1-u)\,\Phi\bigl(\frac{c-\sqrt{\rho}\,q_{1-\alpha}(X)}{\sqrt{1-\rho}}\bigr)},
\end{equation}
and the Basel~I approach. The latter approach just entails the diagonal as
risk contribution curve since it corresponds to purely volume-oriented capital
allocation.

Note that in Figure~\ref{fig:2} the true CVaR curve intersects the diagonal
(Basel~I curve) just at the relative weight $u^\ast$ that corresponds to the
minimum risk portfolio $L(u^\ast)$. The Basel~II curve differs strongly from
the true contribution curve and is completely situated below the diagonal.
This fact could yield the misleading impression that
an arbitrarily high exposure to the additional loan still improves the risk of the
portfolio. However, as the true CVaR curve in Figure~\ref{fig:2} shows, the
diversification effect from pooling with the new loan stops at 5.8\% relative
weight.

To sum up, it can be said that the example shows a shortcoming of the new
Basel II capital requirement rules as they are not sensitive to
concentrations. In addition, the example presents an intuitive bottom-up
approach for calculating contributions that is sensitive to correlation as
well as to concentrations and avoids time-consuming simulations.



\end{document}